\documentclass[twocolumn,aps,prl,showpacs,superscriptaddress,final]{revtex4-1}   
\usepackage{bm}
\usepackage{hyperref}
\hypersetup{
    colorlinks = true,
    linkcolor  = blue,
    citecolor  = blue,
    filecolor  = magenta,
    urlcolor   = cyan
}
\usepackage [english]{babel}
\usepackage [utf8]{inputenc}
\usepackage [T1]{fontenc}
\usepackage[pdftex]{graphicx}
\usepackage{amsmath}
\usepackage{amssymb}
\usepackage{siunitx}
\usepackage{booktabs}
\usepackage{float}
\usepackage{array}
\usepackage{scalerel}

\usepackage[normalem]{ulem}
\usepackage[dvipsnames]{xcolor}
\usepackage[pdftex]{graphicx}
\graphicspath{{./figs},}

\newcommand{\dline}[1]{{D\textsubscript{#1}}}

\newcommand*{\element}[1]{\ensuremath{\mathrm{#1}}}
\newcommand*{\K}{\element{K}}
\newcommand*{\Rb}{\element{Rb}}
\newcommand*{\Cs}{\element{Cs}}

\usepackage{xcolor}

\newcommand{\addressinstphys}{Institute of Physics, University of Freiburg, Hermann-Herder-Str. 3, D-79104 Freiburg, Germany}
\newcommand{\addresseucor}{EUCOR Centre for Quantum Science and Quantum Computing, Albert-Ludwigs-Universität Freiburg, Hermann-Herder-Str. 3, D-79104 Freiburg, Germany}

\begin{document}

\title{Anisotropic fluorescence signals\\ retarded dipole-dipole interactions in a thermal atomic cloud}

\author{Vyacheslav Shatokhin}
\email[]{vyacheslav.shatokhin@physik.uni-freiburg.de}
\affiliation{\addressinstphys}
\affiliation{\addresseucor}

\author{Friedemann Landmesser}
\affiliation{\addressinstphys}

\author{Mario Niebuhr}
\affiliation{\addressinstphys}

\author{Frank Stienkemeier}
\affiliation{\addressinstphys}

\author{Andreas Buchleitner}
\affiliation{\addressinstphys}
\affiliation{\addresseucor}

\author{Lukas Bruder}
\affiliation{\addressinstphys}

\date{\today}

\begin{abstract} 
We experimentally observe and theoretically explain anisotropic 
multiple quantum coherence signals in the fluorescence from dilute 
thermal potassium vapors, at room temperature and 
particle densities $\sim 10^8\ \rm{cm}^{-3}$. We identify the retarded 
part of the geometrically fully resolved inter-atomic, resonant 
dipole-dipole interaction as the crucial ingredient to theoretically 
reproduce all qualitative features of the experimental spectra.
\end{abstract}


\maketitle

{\it Introduction.---} In an atomic gas, spontaneous emission
from an atom's electronic excited states is affected by 
dipole-dipole interactions with surrounding atoms, which 
are mediated by the electromagnetic environment 
\cite{Dicke.1954,Stephen.1964,Hutchinson.1964,Gross.1982}.
In thermal clouds, one is tempted to argue that the 
atoms' 
motion washes out any coherent 
inter-atomic interactions, such that their 
fluorescence properties can be considered that of 
independent particles \cite{Jenkins.2016,Mukamel.2016}.
However, several publications \cite{Dai.2012,Gao.2016,Bruder.2019b,Yu.2019,Yu.2019b, Liang.2021,Landmesser.2023,Bruder.2015} reported multiple quantum coherence (MQC) 
signals, by ultrafast nonlinear spectroscopy methods. 
They oscillate at integer multiples $\kappa \leq 8$ of a single atom's 
dipole transition frequencies, and are observed at room temperature, down to extremely low particle concentrations 
$\sim 8\times 10^6\ \rm{cm}^{-3}$ \cite{Bruder.2019b}.
These observations suggest  coherent interactions 
between at least $\kappa$ atoms, contradict the above 
picture of independent atomic emitters, and triggered 
a hitherto open debate on the origin of the 
experimental observations. Furthermore, they underpin 
evidence for non-trivial quantum optical phenomena 
in thermal many-body systems, with a panoply of 
potential applications for spectroscopic \cite{Pizzey.2022,Fabricant_2023,Uhland_2023} 
and quantum information \cite{Skljarow.2022, Buser.2022,Glorieux_2023,Alaeian_2024} purposes.

In our present contribution, we 
seek to settle 
the above debate, in a joint effort of theory and 
experiment. Combining a full-fledged quantum optical 
treatment of the inter-atomic interaction through the 
atoms' electromagnetic environment with a faithful 
description of the vector-valued character of the atomic 
dipole transitions and some ingredients of disorder 
physics, we identify a directional anisotropy in the 
experimentally recorded $\kappa =1,2$ MQC signal as a 
hallmark of the 
dominant far field character of the coherent 
dipole interaction. This simple observable yields 
unambiguous experimental evidence for our theoretical 
model, in which 
quantum emitters with sufficiently small relative 
velocities remain coupled by the retarded 
dipole-dipole interactions, despite the atoms' thermal 
motion. Our results crucially depend on the accurate account 
of the internal quantum structure of the degenerate 
dipole transitions of alkali atoms. Comparing experimental 
data to our theoretical model's predictions, we can rule 
out the model of independent emitters, as well as the 
hypothesis of only electrostatic dipole-dipole interactions.

\emph{Experimental setup.---} 
\begin{figure}
\includegraphics[width=8.5cm]{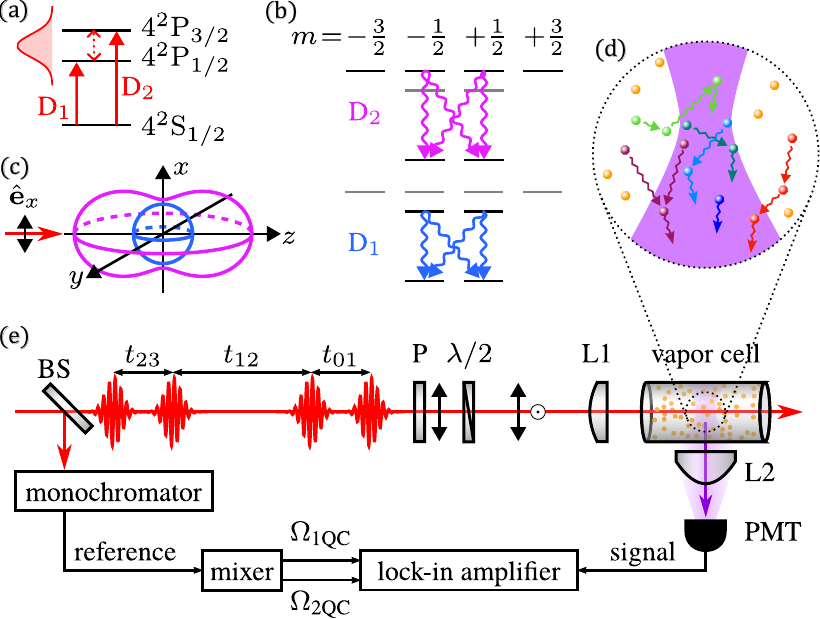}
\caption{\label{fig:setup}%
Main ingredients of the experiment:
(a) Soild red lines indicate the 
K atoms'
\dline{1}, \dline{2}-line transitions \cite{freK}; 
the red dotted arrow the fine structure coherence. (b) Zeeman structure 
of the \dline{1}, \dline{2} transitions, with 
magnetic quantum numbers $m$, and 
fluorescence channels contributing for $\hat{\bf x}$-polarized incoming pulses. (c) 
Associated fluorescence angular distribution; 
details in~\cite{supp}. (d) Disordered, thermal vapor where resonant photon exchange (wiggly arrows) is possible for particles with similar velocities (same color).
(e) Experimental setup (see main text).}
\end{figure}
Most aspects of the experimental setup were described previously\,\cite{Bruder.2019b}. 
Our spectroscopic scheme is based on two-dimensional MQC spectroscopy\,\cite{Yu.2019, Yu.2019b} using the phase 
modulation technique\,\cite{Tekavec.2006, Bruder.2015}. 
Phase-modulated electronic coherences are induced in a potassium (K) vapor (Fig.\,\ref{fig:setup}(a)), at $T\approx\qty{295}{K}$ and particle density $n_0 \approx \qty{5e8}{\per\centi\meter\cubed}$, and are tracked using $M_{\rm cyc}\gg 1$ excitation/detection cycles. 
Figure~\ref{fig:setup}(e) shows our experimental scheme for a single cycle.
Briefly, four linearly polarized Gaussian pulses, each with 
center wavelength $\lambda\approx \qty{768}{\nm}$, energy $\simeq\qty{54}{pJ}$ and duration $\sigma=\qty{190}{\fs}$,  are created at relative times $t_j$ ($j = 0, 1, 2, 3$) in nested Mach-Zehnder interferometers, and phase modulated using acousto-optical modulators, at frequencies $\Omega_j$, in each interferometer arm \cite{Binz:20}. The pulses are focused by 
lens 
L1 into a $L=\qty{5}{\cm}$ long alkali vapor cell. The fluorescence light is 
collected (irrespective of its polarization) by 
lens 
L2, and detected by a photomultiplier tube (PMT), over a solid angle of $\approx \qty{0.38}{\steradian}$. 
The incoming laser polarization with respect to the detection direction (perpendicular to the incoming radiation's wave vector) is controlled with high accuracy by a combination of a linear wire-grid polarizer (P) and a half-wave plate ($\lambda/2$), set either along ($\hat{\bf x}$) or perpendicular ($\hat{\bf y}$) to the 
detection axis. 
A lock-in detection scheme \cite{Tekavec.2006} allows
to extract specific $\kappa$ contributions from the overall 
signal \cite{Bruder.2015}, by scanning the delay time
$\tau:=t_{12} \equiv t_2 - t_1$ of the central pulse pair with a
mechanical delay line, thus recording 
a one-dimensional interferogram
over the $\tau$ axis.
We fix the other delays, $t_{01} = t_{23} = \qty{577}{\fs}$, to avoid possible time-ordering artifacts \cite{Bangert.2023, Hedse.2023} due to the finite pulse duration.
The lock-in reference signal is constructed by detecting the modulated interference of the pulses with a monochromator and filtering it at specific frequencies $\Omega_\text{1QC} = \Omega_{2} - \Omega_{1} = \qty{2.7}{\kHz}$ and $\Omega_\text{2QC} = \Omega_\text{1QC} + \Omega_{3} - \Omega_{0} = \qty{12.7}{\kHz}$ \cite{supp}, producing the 1QC and the 2QC interferograms upon demodulation, respectively.
Finally, a Fourier transform of the interferograms, oscillating 
against $\tau$ at the atomic $D_{1,2}$ transition frequencies 
$\omega_{1,2}$, and at their 
additive combinations $2\omega_{1,2}$ and $\omega_1+\omega_2$, yields the 1QC and 2QC spectra, respectively.

\emph{Theoretical analysis.---} 
The MQC signals are theoretically modeled by a master equation ansatz describing randomly 
oriented atomic dipoles which interact through their common electromagnetic environment 
\cite{Shatokhin.2005,Ames.2022}, where we built in the internal structure of the experimentally 
probed potassium atoms \cite{hyperfine} (cf. Fig.~\ref{fig:setup}(a)).

In accordance with the experimental protocol, we seek 
the total far-field fluorescence intensity emitted into a solid angle $\Omega$ around the 
unit vector $\hat{\bf k}$ pointing towards the detector, 
during one excitation/detection cycle of duration $T_{\rm cyc}$ \cite{Ames.2022}:
\begin{equation}
    I_{\Delta\hat{\bf k}}(T_{\rm cyc};\tau)= \int_0^{T_{\rm cyc}}\!\!\!\!\int {\rm d}t\;{\rm d}\Omega\; \langle {\bf E}^{(-)}_{\hat{\bf k}}(t;\tau)\cdot {\bf E}^{(+)}_{\hat{\bf k}}(t;\tau)\rangle,
    \label{eq:int}
\end{equation}
where ${\bf E}^{(\pm)}_{\hat{\bf k}}(t;\tau)$ is the positive/negative frequency part of the electric field operator, which can be expressed in terms of the atomic dipole operators and implies a summation over all populated polarization channels.
$\langle\ldots\rangle$ denotes the quantum mechanical expectation value \cite{supp}.
The correlator under the integral in Eq.~\eqref{eq:int} is deduced from 
the solution of the master equation for three randomly placed (at 
distances much larger than the wave length associated with the 
atomic transition frequency) and oriented K atoms, each initially 
uniformly populating the ground state's Zeeman sublevels. 
We exploit the separation of the time scales of the problem \cite{Ames.2022}: During the action of the ultrashort pulses, 
radiative decay and 
dipole-dipole interactions are neglected.
Beyond these times, 
pairwise dipole-dipole interactions are treated perturbatively, 
at 
lowest (second) order in the interaction tensor {$\bf T$} (see below), which survives the average over the atoms'
random positions \cite{Ames.2022,ketterer2014,Binninger.2019}. 
Consequently, recurrent scattering  whereupon a photon bounces 
back and forth between the same pair of coupled atoms is neglected \cite{Ames.2022}.  

To account for the pulsed excitation,
we use a simplified scheme with two, instead of four, pulses, which however captures all 
relevant features of the experiment~\cite{supp}. 
The pulses are modeled as delta
pulses with identical 
pulse areas \cite{twopulses}, and their (near-resonant) interaction with the atoms is treated \emph{non-perturbatively} (in the sense of admitting saturation effects).
Laser pulse $j$ ($j=1,2$) imprints the instantaneous phases 
$\varphi_j^{(\alpha)}={\bf k}_{\rm L}\cdot {\bf r}_{\alpha 0}+\omega_1 t_j+\Omega_j \tau_m+\Delta_\alpha t_j$ and $\phi_j^{(\alpha)}={\bf k}_{\rm L}\cdot {\bf r}_{\alpha 0}+\omega_2 t_j+\Omega_j \tau_m+\Delta_\alpha t_j$ upon the dipole transitions \dline{1} and \dline{2} of atom $\alpha$, respectively, with ${\bf r}_{0\alpha}$ the initial position of the atom at the beginning of cycle $m$, $k_L$ the driving field wave vector, 
$\tau_m=mT_{\rm cyc}$, and $\Delta_\alpha={\bf k}_{\rm L}\cdot {\bf v}_\alpha$ the Doppler shift of the atom when moving with velocity ${\bf v}_\alpha$ \cite{maxwell}. 
Furthermore, the spectrally broad pulses induce atomic fine-structure coherence (see red dotted arrow in Fig.~\ref{fig:setup}(a)), whose implication is discussed later. 
After the two pulses, the populations of the $m=\pm 1/2$ sublevels of level $4^{2}\!P_{1/2}$ ($4^{2}\!P_{3/2}$) carry the \emph{position-independent} phase $\varphi_{21}^{(\alpha)}=\varphi_2^{(\alpha)}-\varphi_{1}^{(\alpha)}$ ($\phi_{21}^{(\alpha)}=\phi_2^{(\alpha)}-\phi_{1}^{(\alpha)}$), which 
identifies the 1QC signal
at 
$\omega_1$ ($\omega_2$) \cite{Ames.2021b}. 

To emit 2QC signals, atoms that have interacted with the laser must also interact \emph{twice} with each other \cite{Ames.2022}.
The resonant dipole-dipole interaction Liouvillian is proportional to the $3\!\times\! 3$ tensor ${\bf T}$, with Cartesian components \cite{supp}
\begin{align}
({\bf T})_{kl}&=\frac{3\gamma}{4}\frac{ \exp\{ik_0r_{\alpha\beta}(t)\}}{k_0r_{\alpha\beta}(t)}\left[(\delta_{kl}-(\hat{\bf n})_k(\hat{\bf n})_l)\right.\label{eq:T}\\
&\left.+(\delta_{kl}-3(\hat{\bf n})_k(\hat{\bf n})_l)\left(\frac{i}{k_0r_{\alpha\beta}(t)}-\frac{1}{(k_0r_{\alpha\beta}(t))^2}\right)\right],\nonumber
\end{align}
$r_{\alpha\beta}(t)=|{\bf r}_{0\alpha}-{\bf r}_{0\beta}+({\bf v}_\alpha-{\bf v}_\beta)\;t|$, $\hat{\bf n}$ 
the unit vector connecting atoms $\alpha$ and $\beta$, $\gamma=1/\tau_{\rm spon}$ 
the spontaneous decay rate of either one of the excited states of the 
\dline{1} 
and \dline{2} transitions, and $k_0=(\omega_1+\omega_2)/2c$ 
the mean of the wave numbers associated with the atomic transition frequencies. 
${\bf T}$ quantifies the coupling  amplitude of the field emitted by one dipole to another dipole \cite{LandauII}; it 
mediates the retarded dipolar interaction for a broad range of interatomic separations. While most studies dedicated to MQC signals \cite{Dai.2012,Gao.2016,Lomsadze.2018,Bruder.2019b,Yu.2019,Yu.2019b, Liang.2021} account only for the electrostatic, or near-field, dipolar interactions scaling as $\sim  r_{\alpha\beta}^{-3}$,   
our 
present results are obtained by 
retaining the full
form (2)
of 
${\bf T}$,
thereby
incorporating
the far field interactions. These, 
scaling as $\sim  r_{\alpha\beta}^{-1}$, 
mediate multiple scattering processes \cite{Shatokhin.2005,ketterer2014,Binninger.2019}. 



Robust double scattering contributions to experimental observables 
stem from 
products of 
interaction amplitudes proportional to $({\bf T})_{kl}(t')$ and $({\bf T}^\ast)_{nm}(t'')$, respectively, at distinct times $0<t',t''\sim \tau_{\rm spon}$, 
which survive the disorder average \footnote{In some analogy to robust -- against disorder averaging -- interference contributions which give rise to the coherent backscattering signal upon multiple scattering of light off clouds of cold atoms (see \cite{Shatokhin.2005,Binninger.2019}, and references therein).}. For such terms, the 
exponential phase factors of $({\bf T})_{kl}(t')$ and $({\bf T}^\ast)_{nm}(t'')$ (see eq.~(\ref{eq:T})) must 
approximately compensate each other.
This implies that the atomic velocities of the interacting atoms must obey the inequality $|{\bf v}_\alpha-{\bf v}_\beta|\tau_{\rm spon}\ll \lambda$, which is the condition for the atoms to be in the same \emph{velocity class}.
Although it was 
argued earlier that 2QC signals in a dilute thermal gas originate from interacting atoms which move with near-vanishing 
relative velocity 
\cite{Lomsadze.2018},  
this requirement was 
then inferred from the $r_{\alpha\beta}^{-3}(t)$-dependence of the electrostatic dipolar interactions \cite{Lomsadze.2018}, without account of the indispensable disorder average over the atomic positions.
It is the robustness with respect to the latter which we here identify as the actual cause for 2QC signals to survive.
 Finally, we only retain the {\it resonant} interactions between the  same dipole transitions  of  distinct atoms (that is, \dline{1}-\dline{1} and \dline{2}-\dline{2}); 
 the off-resonant dipolar interactions yield terms that
 oscillate at frequency $\simeq|\omega_1-\omega_2|\approx \qty{1.73}{\THz}$, and 
 effectively average
 out during the cycle duration \cite{puri}.


The fluorescence signal of interacting atoms $\alpha$ and $\beta$ satisfying the above conditions 
 is modulated by phases $\varphi_{21}^{(\alpha)}+\varphi_{21}^{(\beta)}$ and $\phi_{21}^{(\alpha)}+\phi_{21}^{(\beta)}$. Upon demodulation, this gives rise to 2QC signals at $2\omega_1$ and $2\omega_2$, i.e., to the 2\dline{1} and 2\dline{2} resonances. However, this does not explain the previous observations of cross resonances \dline{1}\dline{2}\,\cite{Dai.2012, Bruder.2015, Bruder.2019b, Yu.2019, Yu.2019b}. 
 These peaks emerge due to the interplay between
 fine-structure coherence and dipolar interactions, which 
gives rise to a modulation of the excited state populations by the 
phases  
$\phi_{21}^{(\alpha)}+\varphi_{21}^{(\beta)}$ and $\varphi_{21}^{(\alpha)}+\phi_{21}^{(\beta)}$. 
 These phases then bring about the cross resonance \dline{1}\dline{2}, at $\omega_1+\omega_2$. 
Note that, if only one of the two dipole-dipole interacting atoms has been excited by the laser pulses, a contribution to the respective 1QC signal due to double scattering appears. In general, a $\kappa$QC signal can feature contributions from $s\geq \kappa$ particles, 
combining $2(s-1)$ interaction amplitudes. 

To summarize, the spectra $\tilde{I}_{1,\Delta\hat{\bf k}}(\omega)$ and $\tilde{I}_{2,\Delta\hat{\bf k}}(\omega)$, respectively, of the 1QC and 2QC signals emitted into a solid angle  around $\hat{\bf k}$, can be expanded into a series:
\begin{subequations}
\begin{align}
\tilde{I}_{1,\Delta\hat{\bf k}}(\omega)&=\tilde{I}^{[0]}_{1,\Delta\hat{\bf k}}(\omega)+\tilde{I}^{[2]}_{1,\Delta\hat{\bf k}}(\omega)+\tilde{I}^{[4]}_{1,\Delta\hat{\bf k}}(\omega)+\ldots,\label{eq:I1}\\
\tilde{I}_{2,\Delta\hat{\bf k}}(\omega)&=\tilde{I}^{[2]}_{2,\Delta\hat{\bf k}}(\omega)+\tilde{I}^{[4]}_{2,\Delta\hat{\bf k}}(\omega)+\ldots,\label{eq:I2}
\end{align}
\label{eq:contrib_spectra}
\end{subequations}
where the super-(sub-)scripts indicate the number of 
involved dipole
interaction amplitudes (the 
order $\kappa$ of the MQC 
signal emitted into a solid 
angle around $\hat{\bf k}$). 

\emph{Spectra and anisotropy of MQC signals.---}
\begin{figure}
\includegraphics[width=3.375in]{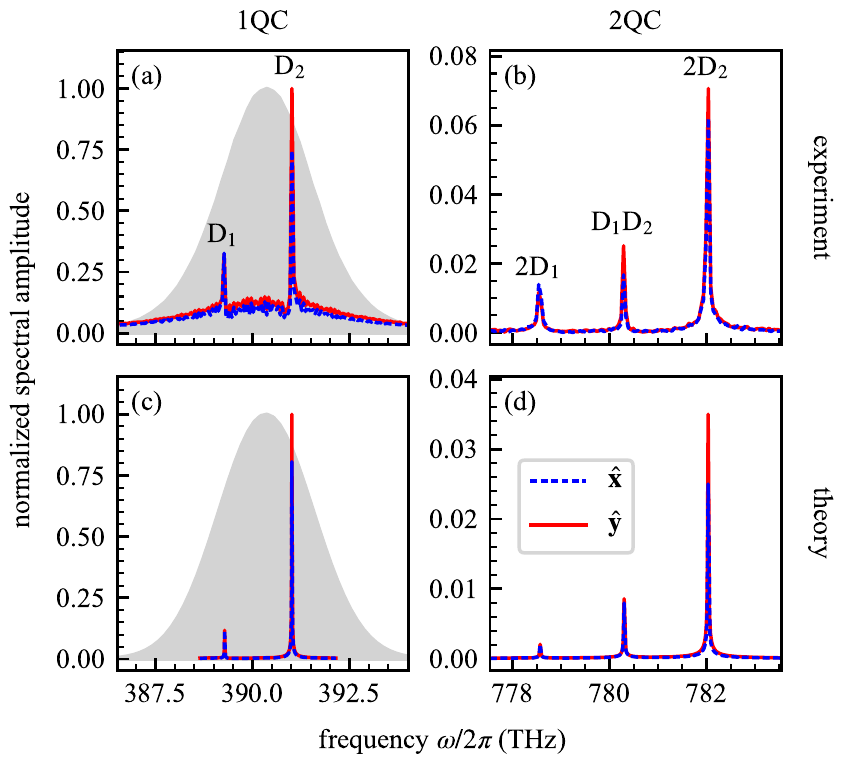}
\caption{\label{fig:spectra}%
Experimental (a,b) 
vs. theoretical (c,d) 1QC and 2QC spectra, for 
detection in the $\hat{\bf x}$- (blue) and $\hat{\bf y}$-direction (red), respectively, with respect to the injected laser polarization (compare Fig.~1(c,e)) and 
spectrum (gray). The broad background in (a) is due to stray light. The broader linewidths in the experiment 
reflect the experimental 
detector response function. The theoretical spectra are rescaled as described in the main text, with  
$N_{\det}=10^9$, $k_0=8.18\times 10^4$ m$^{-1}$, and $\bar{r}=7.5$ mm.
}
\end{figure}
In Fig.~\ref{fig:spectra}(a,b), we present examples of the measured 1QC and 2QC spectra that are normalized by $|\tilde{I}_{1,\Delta\hat{\bf y}}(\omega_2)|$, for detection parallel ($\hat{\bf x}$, blue dashed) and perpendicular ($\hat{\bf y}$, red solid) to the injected polarization direction. 
The experimental spectra exhibit two and three resonances for 1QC and 2QC spectra, respectively, as 
anticipated above; they are 
similar to the previously reported observations \cite{Dai.2012, Bruder.2015, Bruder.2017b, Bruder.2019b, Yu.2019b, Liang.2021}, and are
in good qualitative agreement with our theoretical results (see Fig.~\ref{fig:spectra}(c,d)). In theory and experiment, the \dline{2} and 2\dline{2} resonances are stronger than the \dline{1} and 2\dline{1} resonances, respectively, due to the larger oscillator strength of the \dline{2} transition \cite{Steck.20220728}. However, the 
magnitude
of the computed peak
originating from the \dline{2} transition, relative to that of the 
\dline{1} line, is
several times too strong compared to the measured one.
We attribute this to photon frequency redistribution in a thermal vapor \cite{molisch}, which leads to a stronger attenuation of the field in resonance with the \dline{2} rather than with the \dline{1} transition (due to the larger scattering cross section of the \dline{2} transition \cite{Budker.2008}), but 
is not accounted for by our model.


Let us 
now assess the specific role of the retarded part of the dipole-dipole interaction (2) for the observed 1QC and 2QC signals. First, consider the different amplitudes of the 1QC and 2QC spectra. 
For a quantitative comparison between experiment and theory,
we extrapolate the results of our toy model of three atoms to the bulk vapor.
Given 
the density of the particles and the parameters of the detection system \cite{Landmesser.2024c}, the number of particles in the detection volume is estimated 
as $N_{\rm det}\sim 10^8-10^9$ \cite{Landmesser.2024c}. Using this, we rescale the single, double, and triple scattering terms (that is, with the  superscript $[0]$, $[2]$, and $[4]$, respectively) in Eq.~\eqref{eq:contrib_spectra} by $N_{\rm det}$, $N_{\rm det}^2/(k_0\bar{r})^2$, and $N_{\rm det}^3/(k_0\bar{r})^4$, respectively, where $\bar{r}=0.554 n_{\rm vc}^{-1/3}\sim 0.6-1.2$ cm is the average distance between the particles belonging to the same velocity class, estimated from their density $n_{\rm vc}$ \cite{supp,Chandrasekhar1943}.
The thus obtained spectra, normalized by the rescaled value of $|\tilde{I}_{1,\Delta\hat{\bf y}}(\omega_2)|$, are plotted in Fig.~\ref{fig:spectra}(c,d). 
The relative 1QC/2QC magnitudes of the \dline{2} and 2\dline{2} peaks retrieved from the calculations and measurements are $\approx 28$ and $\approx 13$, respectively. 
Note that this result clearly improves over that obtained when only the static (i.e., non-retarded) part of the dipole-dipole interaction is accounted for, leading to an overestimation of the 1QC/2QC amplitude ratio by a factor $10^7$ \cite{Bruder.2019b}, with respect to the values here obtained.


We 
remark that the relatively strong 2QC signals cannot be explained by semiclassical theory of multiple scattering \cite{Lagendijk1996143}, which describes the propagation of a photon as a random walk. One of the key parameters describing the walk is the photon scattering mean free path $\ell$.
For the considered vapor densities, $\ell \gtrsim 1\;{\rm m}\gg L=5\ \rm cm$ (see above) \cite{Weis.2011}, such that photons should travel quasi-ballistically across so optically thin vapor, with, at most, one scattering event \cite{Mercadier:2009pp}. Hence, 2QC signals should not be observed, else their magnitude should be negligible compared to 1QC signals. This contradicts our observations, and it is known that the random-walk model ignores collective effects \cite{Sokolov:19}, nor can it accurately describe the fluorescence emission by thermal quantum scatterers in the transverse direction, while the coupled-dipole model succeeds in this task \cite{Bromley:2016kx}.

\begin{figure}
\includegraphics[width=3.375in]{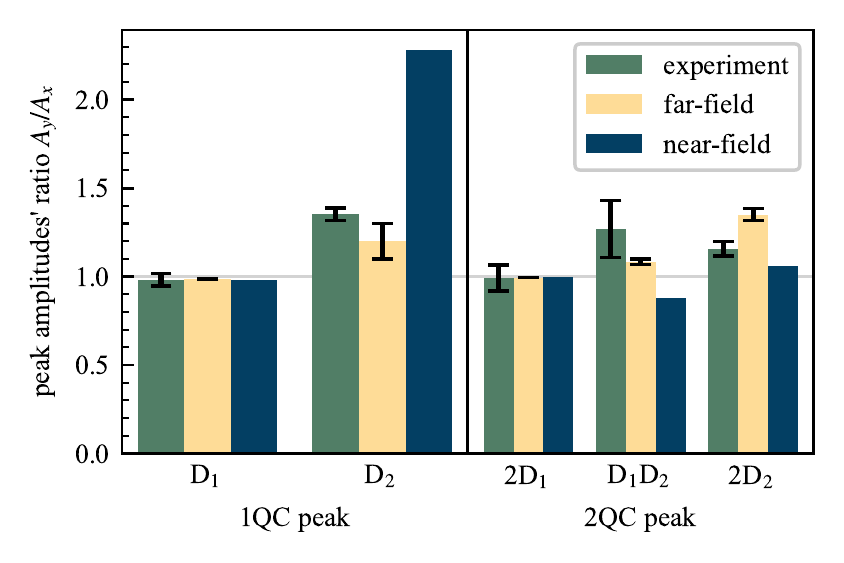}
\caption{\label{fig:ratios}%
Anisotropy of experimental vs. theoretical 1QC and 2QC signals, the latter derived either from the electrostatic near-field contribution to the dipole-dipole interaction (2) alone, or accounting also for the far-field  
contributions.
Error bars indicate 
standard deviations for the experimental record,
and uncertainties of $N_{\rm det}$ and $\bar{r}$, for theory (see main text).
}
\end{figure}

Another feature of the MQC signals 
characteristic of both, the measured and the calculated spectra, 
are the (un)equal amplitudes of the spectral peaks $A_x$ and $A_y$ associated with the \dline{1} (\dline{2}) transitions in $\hat{\bf x}$ and $\hat{\bf y}$ directions (see Fig.~\ref{fig:spectra}). 
We examined this (an)isotropy further by looking at the peak ratios along the two observation directions $\hat{\bf x}$ and $\hat{\bf y}$.
The bars of the histogram in Fig.~\ref{fig:ratios} present the experimental and theoretical values of the peak ratios, for the five peaks exhibited jointly by 1QC and 2QC spectra. 
For each peak, we furnish two theoretical bars: one evaluated using the full dipole interaction (2), including the 
far-field 
contributions, and the other one 
retaining solely 
the 
electrostatic interactions, which hitherto
were 
almost exclusively used  in theoretical treatments of MQC signals in alkali atom vapors.

As expected, the \dline{1} and 2\dline{1} peaks (see Fig.~\ref{fig:ratios}) 
exhibit isotropy of the spontaneous emission ($A_y/A_x\approx 1$) from the excited state's Zeeman sublevels which are equally populated by the linearly polarized pulses (see Fig.~\ref{fig:setup}(b,c) and\,\cite{supp}). 
Conversely, peaks involving \dline{2} transitions are anisotropic.  
The computed 1QC \dline{2} signal exhibits the ratio $A_y/A_x=1.2(1)$, which  
is in reasonably good agreement with the experimental observation ($A_y/A_x=1.35(4)$). These values are significantly smaller than the ratio $A_y/A_x\approx 2.25$ characteristic of independent atoms \cite{supp}, or than  
the value $A_y/A_x\approx 2.28$ derived by us
for 
a system  of atoms coupled by the near-field interaction alone.  
Thus, the reduced anisotropy of 1QC signals stems from the redistribution of Zeeman level populations mediated by the far-field dipole-dipole interactions. This 
conclusion is further corroborated by the anisotropy of 2QC signals involving the \dline{2} transition. The far-field dipolar 
contributions yield 
signal ratios $A_y/A_x>1$ of the 2\dline{2} and \dline{1}\dline{2} peaks, respectively, in qualitative agreement with experiment, while the electrostatic interactions alone, in contrast, result invert  
this ratio, for the \dline{1}\dline{2} peak (see Fig.~\ref{fig:ratios}).
Finally, we note that population redistribution may also be mediated by thermal atomic collisions. However, in our many-body system, interatomic collisions occur on 
a time scale $\tau_{\rm coll}\sim \qty{1.5}{\ms}\gg \tau_{\rm spon}\approx \qty{26}{\ns}$ \cite{supp}, and are thus not relevant for the MQC signals. 

\emph{Conclusion.---} 
Our experimental results and theoretical analysis of the amplitude ratio between 1QC and 2QC signals and their anisotropy in very dilute thermal vapors of alkali atoms 
provide
clear evidence of the far-field dipole-dipole interactions 
as the physical mechanism underlying MQC signals. 
The interatomic interactions persist for quantum emitters moving very slowly relative to each other, 
and the MQC detection scheme 
extracts their signatures despite strong inhomogeneous broadening. 
The ability of this technique to detect weak interactions without the need for confined geometries \cite{Skljarow.2022,Alaeian_2024} paves a promising way towards the sensing and control of dilute thermal systems.


\begin{acknowledgments}
We thank U. Bangert and I.M. Sokolov for insightful discussions.
This work was supported
by the European Research Council (ERC) with the Advanced Grant COCONIS (694965),
by the German Research Foundation (DFG; IRTG 2079, RTG 2717), 
and by the state of Baden-Württemberg through bwHPC. V. S. and
A. B. acknowledge partial funding and support through the
Strategiefonds der Albert-Ludwigs-Universität Freiburg 
and the Georg H. Endress Stiftung.
\end{acknowledgments}


\bibliography{sources}

\begin{thebibliography}{55}
\bibitem [{fre()}]{freK}%
  \BibitemOpen
  \href@noop {} {\bibinfo  {journal} {Resonance frequencies of $4^2S_{1/2}\rightarrow 4^2P_{1/2}$ and $4^2S_{1/2}\rightarrow 4^2P_{3/2}$ transitions are $\omega_1=2\pi\times\qty{389.6}{\THz}$ (\dline{1}) and $\omega_2=2\pi\times\qty{391.3}{\THz}$ (\dline{2})}\ }\BibitemShut {NoStop}%
\bibitem [{hyp()}]{hyperfine}%
  \BibitemOpen
  \href@noop {} {\bibinfo  {journal} {The hyperfine structure is neither resolved in the experiment nor included in our theoretical description}\ }\BibitemShut {NoStop}%
\bibitem [{two()}]{twopulses}%
  \BibitemOpen
  \href@noop {} {\bibinfo  {journal} {The results of the two pulse excitation are equivalent to the four pulse results, upon rescaling \cite{supp}}\ }\BibitemShut {NoStop}%
\bibitem [{max()}]{maxwell}%
  \BibitemOpen
\bibfield  {journal} {  }\href@noop {} {\bibinfo  {journal} {Our results are averaged over the Maxwell-Boltzmann distribution of $\Delta_\alpha$}\ }\BibitemShut {NoStop}%
\end{thebibliography}

\clearpage

\onecolumngrid

\section{
Supplemental Material
}
\noindent
\subsection{Accumulation effects}
In the experiment, the time interval $T_\mathrm{cyc}\approx 10$ ns between consecutive laser cycles is of the order of the natural lifetime $\tau_{\rm spon}\approx 26$ ns, which in principle can lead to the accumulation of spurious nonlinear signals over several laser cycles\,\cite{osipov_accumulation_2017}. 
We overcome this problem 
by employing four laser pulses, where each pulse interacts only once with the vapor. 
Thereby, the system remains in a coherent superposition state during each of the time intervals $t_{01},\,t_{12},\,t_{23}$ (see the main text). 
Hence, contributions involving multiple laser cycles are not detected in the four-pulse scheme, since the (multi)atomic coherences decay significantly faster (on a time scale $< 1\,$ns) than populations, due to the ensemble inhomogeneity. 

An analogous detection scheme can be implemented with only two laser pulses which interact multiple times with the system\,\cite{Bruder.2015}. 
However, in this case the time interval $T_\mathrm{cyc}$ between consecutive laser cycles should be taken $\gg \tau_{\rm spon}$, which ensures the absence of artifacts from accumulation over multiple laser cycles. 
Thus, a two-pulse scheme with $T_{\rm cyc}\gg \tau_{\rm spon}$ is well justified for the theoretical description of the experiment. Furthermore, the results of the two excitations can be mapped onto those of the four-pulse excitation, upon a simple rescaling. Indeed, we have checked that for 1QC signals, the two-pulse excitation produces the same peak amplitudes as the four-pulse excitation. However, for 2QC signals, only the amplitude of the cross peak in the two- and four-pulse excitation setup is the same, whereas the amplitudes of the 2\dline{1} and 2\dline{2} peaks in the four-pulse scheme are twice as large. The difference stems from the larger number of processes that contribute to the 2\dline{1} and 2\dline{2} peaks in the four-pulse excitation scheme (see Table \ref{tab:phases}). Therefore, the results of the two-pulse excitation coincide with those of the four-pulse excitation, after multiplying by two the amplitudes of the 2\dline{1} and 2\dline{2} signals.
\begin{table}[h!]
\caption{Peaks of the 2QC spectra and the relevant modulation phases in the case of the two- and four-pulse excitation. The phases are introduced in the main text, see also Eq.~\eqref{sm:Ll} below.}
\label{tab:phases}
	\centering
	\begin{tabular}{ccc}
\hline\hline 
		Peak & 	\multicolumn{2}{c}{Modulation phases} \\
	  &	2 pulses&4 pulses\\
		\hline\\
 \vspace{3pt}
 2\dline{1} &\hspace{5pt} $\varphi_{21}^{(\alpha)}+\varphi_{21}^{(\beta)}$ &\hspace{5pt}$\varphi_{30}^{(\alpha)}+\varphi_{21}^{(\beta)}$, $\varphi_{30}^{(\beta)}+\varphi_{21}^{(\alpha)}$\\
2\dline{2}&\hspace{5pt}  $\phi_{21}^{(\alpha)}+\phi_{21}^{(\beta)}$&$\phi_{30}^{(\alpha)}+\phi_{21}^{(\beta)}$, $\phi_{30}^{(\beta)}+\phi_{21}^{(\alpha)}$\\
\dline{1}\dline{2}&\hspace{5pt} $\varphi_{21}^{(\alpha)}+\phi_{21}^{(\beta)}$, $\phi_{21}^{(\alpha)}+\varphi_{21}^{(\beta)}$&$\varphi_{30}^{(\alpha)}+\phi_{21}^{(\beta)}$, $\varphi_{30}^{(\beta)}+\phi_{21}^{(\alpha)}$\\

\vspace{3pt}

\\
		\hline\hline
	\end{tabular}

\end{table}

\subsection{Fluorescence anisotropy}
\begin{figure}
\includegraphics[width=0.7\linewidth]{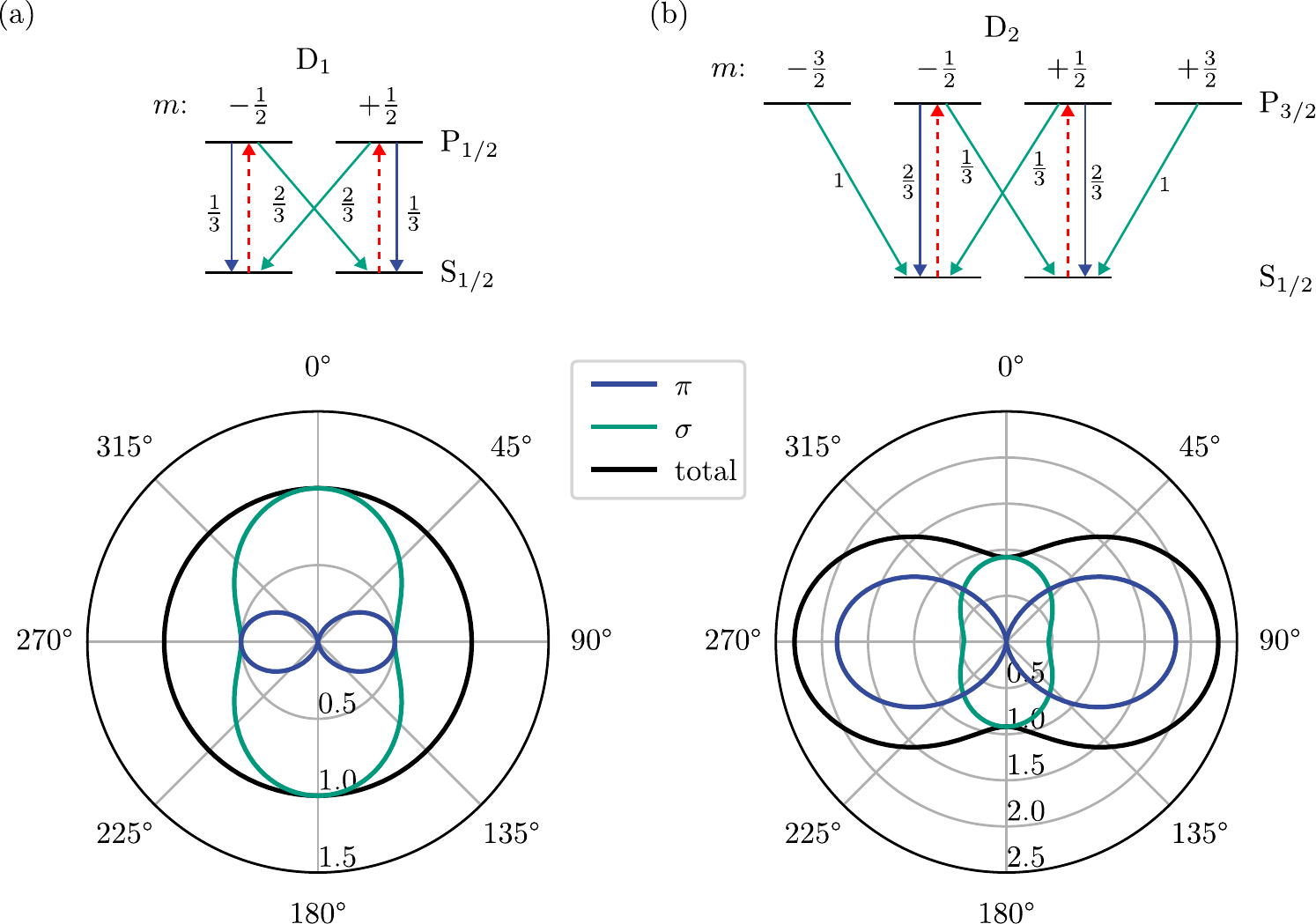}
\caption{\label{fig:anisotropyscheme}%
Angular distribution of fluorescence emission from an isolated alkali atom via the \dline{1} (a) and \dline{2} (b) transitions. Top: Fine structure level scheme 
with degenerate magnetic sublevels ($m$) and the squared Clebsch–Gordan 
coefficients for each transition. 
Dashed red lines indicate the linearly polarized laser excitation, blue arrows the $\pi$ emission and green arrows the $\sigma$ emission, respectively. 
Bottom: The corresponding angular intensity distribution w.r.t. the light’s polarization direction x (0$^\circ$). The total intensity (black) corresponds to the sum of the 
two polarization contributions $\pi$ (blue) and $\sigma$ (green). 
}
\end{figure}
Figure \,\ref{fig:anisotropyscheme} provides an intuitive explanation for the observed anisotropy in the fluorescence emission. 
The atoms are excited by a sequence of laser pulses linearly polarized along the $\hat{\bf x}$-axis, inducing $\pi$-transitions ($\Delta m =0$) in the atoms (red arrows). 
Fluorescence emission, which occurs through atomic $\pi$- and $\sigma$-transitions ($\Delta m=\pm 1$), exhibits an angular distribution of intensity of linearly and circularly polarized light characterized by the corresponding Clebsch-Gordon coefficients (see Fig.~\ref{fig:anisotropyscheme}). The total intensity, summed over all polarizations,  
is isotropic for the \dline{1} line and anisotropic for the \dline{2} line. 
Note that for an isolated atom, the sublevels with $m=\pm 3/2$ will not be excited by linear polarized light. 
However, the dipolar interactions  between the atoms can lead to excitation of these sublevels, which will effectively reduce the anisotropy of the \dline{2} and 2\dline{2} signals, as we observed in the experiment. 

\subsection{Theoretical description}
\emph{Spontaneous emission of alkali atoms. ---}
The positive frequency part of the far-field electric field operator of the light spontaneously emitted by atoms in Eq.~(1) of the main text is given by the expression \cite{Ames.2022}
\begin{equation}
    {\bf E}_{\hat{\bf k}}^{(+)}(t)\approx\frac{\omega_0 d}{4\pi \epsilon_0c^2 R}\sum_{\alpha=1}^N
    \left\{{\bf  D}_{\alpha}(t)-\hat{\bf k}[\hat{\bf k}\cdot{\bf D}_{\alpha}(t)]\right\}e^{-i {\bf k}\cdot {\bf r}_\alpha},
    \label{sm:E+}
\end{equation}
where ${\bf D}_{\alpha}(t)=({\bf D}_{\frac{1}{2},\alpha}+{\bf D}_{\frac{3}{2},\alpha})$ is the lowering part of the vector atomic dipole operator $\boldsymbol{\cal  D}_\alpha(t)=({\bf D}_{\alpha}(t)+{\bf D}^\dagger_{\alpha}(t))$,
and ${\bf D}_{\frac{1}{2},\alpha}$ (${\bf D}_{\frac{3}{2},\alpha}$) is the lowering operator associated with the transition \dline{1} (\dline{2}); scalar $d$ is expressed through the reduced matrix elements as $d\equiv \langle J_{\frac{1}{2}}||{\bf D}_{\frac{1}{2}}||J_{\frac{1}{2}}\rangle/\sqrt{2}\approx \langle J_{\frac{1}{2}}||{\bf D}_{\frac{3}{2}}||J_{\frac{3}{2}}\rangle/2$ \cite{Steck.20220728},  $\omega_0=(\omega_1+\omega_2)/2$, $c$ is the speed of light, and $R$ is the distance from each atom to the detector. The prefactor $\omega_0d/(4 \pi \epsilon_0 c^2 R)$ ignores the difference between the transition frequencies of the \dline{1} and \dline{2} transitions (however, this difference is crucial for, and is retained in, the phase factors $\varphi_j^{(\alpha)}$ and $\phi_j^{(\alpha)}$ defined in the main text) and the variation of the distances from individual atoms to the detector.   
The lowering operator ${\bf D}_{J_{\rm e}}$ of a transition with the total angular momentum of the excited state $J_{\rm e}$ reads (we drop index $\alpha$ for brevity)
\begin{equation}
{\bf D}_{J_{\rm e}}=\sum_{q=-1}^{+1}\sum_{m_g=-J_{\rm g}}^{J_{\rm g}}(-1)^q\hat{\bf e}_q\langle J_{\rm g} m_q,1q|J_{\rm e} m_{\rm g}+q\rangle |J_{\rm g} m_{\rm g}\rangle\langle J_{\rm e} m_{\rm g}+q|,
\label{sm:Dj}
\end{equation}
where $\hat{\bf e}_q$ is the unit vector in the spherical basis and $\langle J_{\rm g}m_q,1q|J_{\rm e}m_{\rm g}+q\rangle$ is the Clebsch-Gordan coefficient. 

The fluorescence intensity is proportional to $\langle  {\bf E}_{\hat{\bf k}}^{(-)}(t) {\bf E}_{\hat{\bf k}}^{(+)}(t)\rangle$ and, by Eq.~\eqref{sm:E+}, depends on $N^2$ two-atom correlation functions. However, the correlators of different atoms contain position-dependent phases $e^{\pm i {\bf k}\cdot( {\bf r}_\alpha(t)-{\bf r }_\beta(t))}$. Such terms do not survive the average over the atomic coordinates that needs to be performed in our disordered system; therefore, we drop all terms containing position-depending phases of different atoms. We thus arrive at the expression which includes a single sum over atomic correlators of all particles \cite{Ames.2021b,Ames.2022}:
\begin{equation}
   \langle  {\bf E}_{\hat{\bf k}}^{(-)}(t) {\bf E}_{\hat{\bf k}}^{(+)}(t)\rangle=f^2\sum_{J_{\rm e}=\frac{1}{2},\frac{3}{2}}\sum_{\alpha=1}^N\langle {\bf D}^+_{J_{\rm e},\alpha}(t^\prime)\cdot({\bf I}-\hat{\bf k}\hat{\bf k})\cdot{\bf D}_{J_{\rm e},\alpha}(t)\rangle,
   \label{sm:corr}
\end{equation}
where $f^2=(\omega_0d)^2/(4\pi \epsilon_0 c^2 R)^2$, ${\bf I}$ is the unit dyadic. Integration of Eq.~\eqref{sm:corr} over the wave vectors $\hat{\bf k}$ within the solid angle subtended by the detection system and over the fluorescence detection time $t$ yields Eq. (1) of the main text. This expression still depends on position-dependent phases of the same atoms. In the final result, we drop these terms and retain only those terms that do not include any position-dependent phases. These terms yield contributions that are robust under the disorder average \cite{Ames.2022}. We selected these contributions from the general analytical expression for the fluorescence intensity \cite{Ames.2022} (also see below). 

\emph{Master equation. ---}
For the assessment of the atomic correlators in \eqref{sm:corr}, we used a master equation for the quantum mechanical expectation value of an atomic operator $Q$ \cite{Ames.2022}. In the interaction picture with respect to the Liouvillian of free atoms, ${\cal L}_0=\sum_{\alpha=1}^N{\cal L}_0^\alpha$, the master equation reads \cite{Ames.2022}:
\begin{equation}
\langle\dot{Q}\rangle=\langle({\cal L}_L+{\cal L}_\gamma+{\cal L}_{\rm int})Q\rangle,
\label{sm:meq}
\end{equation}
where ${\cal L}_L=\sum_{\alpha=1}^N{\cal L}_L^\alpha$, ${\cal L}_\gamma=\sum_{\alpha=1}^N{\cal L}_\gamma^\alpha$ and ${\cal L}_{\rm int}=\sum_{\alpha\neq \beta=1}^N{\cal L}_{\alpha\beta}$ are the Liouvillian operators governing the laser-atom interaction, relaxation, and dipole-dipole interactions, respectively. 
The Liouvillians under the sums read
\begin{align}
{\cal L}_L^\alpha Q&=-\frac{i\vartheta_0}{2}[{\bf D}_{\frac{1}{2},\alpha}\cdot\hat{\bf e}^*_{\rm L}e^{-i\varphi^{(\alpha)}_j t_j}+{\bf D}_{\frac{3}{2},\alpha}\cdot\hat{\bf e}_{\rm L}e^{-i\phi^{(\alpha)}_j t_j}+{\rm h.c.}\;,Q],\label{sm:Ll}\\
{\cal L}_\gamma^\alpha Q&=\frac{\gamma}{2}\left({\bf D}_{\frac{1}{2},\alpha}^+\cdot[Q, {\bf D}_{\frac{1}{2},\alpha}]+[{\bf D}_{\frac{1}{2},\alpha}^+,Q]\cdot{\bf D}_{\frac{1}{2},\alpha}]\right)+\left(\frac{1}{2}\rightarrow \frac{3}{2}\right),\label{sm:Lg}\\
{\cal L}_{\alpha\beta}Q&=\left({\bf D}_{\frac{1}{2},\alpha}^+\cdot{\bf T}{}^*(r_{\alpha\beta}(t),\hat{\bf n})\cdot[Q, {\bf D}_{\frac{1}{2},\beta}]+[{\bf D}_{\frac{1}{2},\beta}^+,Q]\cdot{\bf T}(r_{\alpha\beta}(t),\hat{\bf n})\cdot{\bf D}_{\frac{1}{2},\alpha}\right)+\left(\frac{1}{2}\rightarrow \frac{3}{2}\right),\label{sm:Ldd}
\end{align}
where the atomic dipole operators ${\bf D}_{J_e}$ ($J_e=\frac{1}{2},\frac{3}{2}$) are given by Eq.~\eqref{sm:Dj}, and tensor ${\bf T}$ is defined in Eq.~(2) of the main text. 
Equation \eqref{sm:Ll} describes the Liouvillian of a laser pulse acting upon atom $\alpha$ at time $t_j$. As mentioned in the main text, we modeled pulses by delta-functions multiplied by the pulse areas. We assume that the carrier frequency of the laser pulses $\omega_L=\omega_0$, such that the resonant field amplitude is the same for transitions \dline{1} and \dline{2}. Hence, we apply the same pulse area $\vartheta_0$ for both transitions, while keeping track of the phases $\varphi_j^{(\alpha)}$  and $\phi_j^{(\alpha)}$ imprinted on atomic transitions by the pulses (see the main text). During the action of laser pulses, we ignore the Liouvillians ${\cal L}_\gamma$ and ${\cal L}_{\rm int}$ and evaluate the unitary rotation of an operator $Q$ by ${\cal L}_L^\alpha$ non-perturbatively \cite{Ames.2022}. However, for the pulses' parameters used in the experiment \cite{Landmesser.2024c}(see also the main text)
\begin{equation}
  \vartheta_0\simeq \frac{d}{\hbar}\sqrt{\frac{2 I_L}{c \epsilon_0}}\sqrt{\frac{\pi}{2 \log(2)}}\sigma\simeq 0.3, 
\end{equation}
where $I_L\approx \qty{3.5}{\MW\cdot \cm^{-2}}$ \cite{Landmesser.2024c} and $d=\qty{2.46e-29}{\C\cdot\m}$ \cite{Steck.20220728}.
In this weak driving regime, a perturbative treatment provides quite accurate results \cite{Ames.2021b}. 

In Eqs.~\eqref{sm:Lg}, \eqref{sm:Ldd}, we ignore the small differences (about 1\% \cite{Steck.20220728}) in the spontaneous decay rates of the excited states $4 ^2\!P_{1/2}$ and $4 ^2\!P_{3/2}$, using a single rate $\gamma$. Furthermore, note that both, the radiative relaxation Liouvillian ${\cal L}_\gamma^\alpha$ and the dipolar interactions Liouvillian ${\cal L}_{\alpha\beta}$ features only resonant terms (that is, \dline{1}-\dline{1} and \dline{2}-\dline{2} transitions are coupled). The terms describing the cross coupling \dline{1}-\dline{2} are not included in our Liouvillians; these terms oscillate at frequency $\omega_2-\omega_1\approx \qty{1.73}{\THz}$ and are effectively averaged out during one cycle  \cite{puri}.

We consider $N=3$ potassium atoms located in the far-field ($k_0r_{\alpha\beta}\gg 1$) of each other. We assume that only atom $\alpha$ and/or $\beta$ can interact with the laser field; the third atom is not driven by the laser, but participates in the dipole-dipole interactions. The  presence of the third atom allows to improve the agreement with experiment  for the anisotropy of 2QC signals, as compared to the two-atom case. We exploit the separation of time scales of the problem \cite{Ames.2022} and seek time-dependent solutions of Eq.~\eqref{sm:meq} perturbatively up to fourth order in the coupling strength $|{\bf T}|$, using the full form of tensor ${\bf T}$. In contrast, the near-field results are derived by retaining only the terms $\sim (k_0r_{\alpha\beta})^{-3}$.

The results obtained for fixed configurations of atoms contain a tremendous number of terms, but most of them carry position-dependent phases and, hence, they do not survive the disorder average. Robust contributions are selected using the criteria outlined in the main text and Refs.~\cite{Ames.2021b,Ames.2022}. 
The angular averages are carried out as described in Refs.~\cite{Ames.2022,ketterer2014}.

The average time-dependent fluorescence intensity emitted in direction $\hat{\bf k}$   
admits the following expansion: 
\begin{align}
    I_{\hat{\bf k}}(t,\tau)&=\sum_{l=-1}^{+1}\left[(I^1_{\frac{1}{2}})^{[0]}_{l,\hat{\bf k}}(t,\tau)e^{i\varphi^{(\alpha)}_{21}l}+(I^1_{\frac{1}{2}})^{[2]}_{l,\hat{\bf k}}(t,\tau)e^{i\varphi^{(\alpha)}_{21}l}+(I^1_{\frac{1}{2}})^{[4]}_{l,\hat{\bf k}}(t,\tau)e^{i\varphi^{(\alpha)}_{21}l}\right.\nonumber\\
   &\left.+(I^1_{\frac{3}{2}})^{[0]}_{l,\hat{\bf k}}(t,\tau)e^{i\phi^{(\alpha)}_{21}l}+(I^1_{\frac{3}{2}})^{[2]}_{l,\hat{\bf k}}(t,\tau)e^{i\phi^{(\alpha)}_{21}l}+(I^1_{\frac{3}{2}})^{[4]}_{l,\hat{\bf k}}(t,\tau)e^{i\phi^{(\alpha)}_{21}l}\right.\nonumber\\
    &\left.+(I^2_{\frac{1}{2}})^{[2]}_{l,\hat{\bf k}}(t,\tau)e^{i(\varphi^{(\alpha)}_{21}+\varphi^{(\beta)}_{21})l}
    +(I^2_{\frac{1}{2}})^{[4]}_{l,\hat{\bf k}}(t,\tau)e^{i(\varphi^{(\alpha)}_{21}+\varphi^{(\beta)}_{21})l}\right.
    \nonumber\\ 
&\left.+(I^2_{\frac{3}{2}})^{[2]}_{l,\hat{\bf k}}(t,\tau)e^{i(\phi^{(\alpha)}_{21}+\phi^{(\beta)}_{21})l}
    +(I^2_{\frac{3}{2}})^{[4]}_{l,\hat{\bf k}}(t,\tau)e^{i(\phi^{(\alpha)}_{21}+\phi^{(\beta)}_{21})l}\right.
    \nonumber\\   
    &\left.+(I^2_{\frac{1}{2},\frac{3}{2}})^{[2]}_{l,\hat{\bf k}}(t,\tau)e^{i(\varphi^{(\alpha)}_{21}+\phi^{(\beta)}_{21})l}
    +(I^2_{\frac{1}{2},\frac{3}{2}})^{[2]}_{l,\hat{\bf k}}(t,\tau)e^{i(\phi^{(\alpha)}_{21}+\varphi^{(\beta)}_{21})l}\right.\nonumber\\
  &\left.+(I^2_{\frac{1}{2},\frac{3}{2}})^{[4]}_{l,\hat{\bf k}}(t,\tau)e^{i(\varphi^{(\alpha)}_{21}+\phi^{(\beta)}_{21})l}
    +(I^2_{\frac{1}{2},\frac{3}{2}})^{[4]}_{l,\hat{\bf k}}(t,\tau)e^{i(\phi^{(\alpha)}_{21}+\varphi^{(\beta)}_{21})l}\right],  
    \label{sm:I}
\end{align}
where $(I^{\kappa}_{J_e})^{[2(s-1)]}_{l,\hat{\bf k}}(t,\tau)$ and $(I^{\kappa}_{J_e,J'_e})^{[2(s-1)]}_{l,\hat{\bf k}}(t,\tau)$ are contributions to $\kappa$\dline{1}, $\kappa$\dline{2}  peaks (terms with $J_e=\frac{1}{2}$ or $J_e= \frac{3}{2}$) or to the cross peak \dline{1}\dline{2} (terms with $J_e,J'_e=\frac{1}{2},\frac{3}{2}$) stemming from $s=1,2,3$ atoms. The 1QC and 2QC spectra as given in the main text by Eqs. (2a) and (2b), respectively, are obtained via the demodulation procedure, which is compactly expressed by the formula \cite{Ames.2022}
\begin{equation}
\tilde{I}_{\kappa,\Delta\hat{\bf k}}(\omega)=\lim_{{\cal F}\to 0}\int_0^\infty {\rm d}\tau_m\; e^{-({\cal F}+i\kappa \Omega_{21})\tau_m}\int_0^\infty {\rm d}\tau\; e^{-i\omega \tau}\int_0^{T_{cyc}}{\rm d}t\;\int{\rm d}\Omega\; I_{\hat{\bf k}}(t,\tau),
\label{sm:spectra}
\end{equation}
where $\Omega_{21}=\Omega_2-\Omega_1$, ${\cal F}$ determines the filter bandwidth to extract the $\Omega_{21}$-modulation, and we set $T_{\rm cyc}\to \infty$ in the calculations (see the justification above).

\emph{Averaging the spectra over the Doppler distribution. ---}
We performed the integrations in Eq.~\eqref{sm:spectra} analytically, and obtained that the 1QC and 2QC spectra represent Lorentzian peaks. For the 1QC spectra, the two peaks are centered at frequencies $\omega_1+\Delta_\alpha$ and $\omega_2+\Delta_\alpha$; for the 2QC spectra, the peaks are located at frequencies $2(\omega_1+\Delta_\alpha)$,  $2(\omega_2+\Delta_\alpha)$, and $\omega_1+\omega_2+2\Delta_\alpha$ (although up to 3 atoms contribute to the spectra, they must belong to the same velocity class in order to yield robust contributions, thus, their Doppler shifts $\Delta_\alpha$ are the same to a good approximation).
We outline how we perform the average of $\Delta_\alpha$, by considering only the $2$\dline{2} peak (the procedure is the same for the other peaks). The corresponding spectral function reads
\begin{equation}
S(\omega,\Delta_\alpha)=\frac{A}{\gamma +i(\omega-2(\omega_2+\Delta_\alpha))},
\label{sm:S}
\end{equation}
where $A$ is a numerical factor (which varies depending on the observation direction and the resonance frequency), and all other quantities are defined in the main text.
Integrating \eqref{sm:S} over the Maxwell-Boltzmann distribution of $\Delta_\alpha$, $p(\Delta_\alpha)=\exp(-\Delta_\alpha^2/2\bar{\Delta}^2)/(\bar{\Delta}\sqrt{2\pi})$ (here, $\bar{\Delta}$ is the root-mean-square Doppler shift $\approx 770$ MHz) we obtain
\begin{equation}
 \int_{-\infty}^{\infty}\;{\rm d}\Delta_\alpha S(\omega,\Delta_\alpha) p(\Delta_\alpha) =\frac{A\sqrt{\pi}}{4\sqrt{2}\bar{\Delta}}w(z),
\end{equation}
where $w(z)=\exp(-z^2){\rm erfc}(-i z)$ is the Faddeeva function \cite{Happer}, with ${\rm erfc}(x)$ the complementary error function, and $z=(i \gamma -\omega+2\omega_2)/(2\sqrt{2}\bar{\Delta})$. 

\emph{Mean distance $\bar{r}$ between atoms from the same velocity class. ---}
To estimate $\bar{r}$, we use the formula \cite{Chandrasekhar1943}
\begin{equation}
    \bar{r}\approx0.554\;n_{\rm vc}^{-1/3},
    \label{sm:r}
\end{equation}
where $n_{\rm vc}$ is the particle density in the same velocity class. From the main text, the velocities of such particles obey the condition
\begin{equation}
    |{\bf v}_\alpha-{\bf v}_\beta|\sim \sqrt{3}\Delta v\ll \lambda/\tau_{\rm spon},
\end{equation}
where $\Delta v$ is the absolute value of the difference between one velocity component (assumed to be equal for each of the three Cartesian components, hence the prefactor $\sqrt{3}$). The probability to find two particles whose velocity difference lies in the range $(-\Delta v,\Delta v)$ is given by
\begin{equation}
    p(\Delta v)=\frac{1}{\sqrt{2 \pi}\bar{v}}\int_{-\Delta v}^{\Delta v}{\rm d}x \exp(-x^2/(2 \bar{v}^2)).
\end{equation}
Using $\bar{v}\approx 247$ m/s, $\lambda=767$ nm, $\tau_{\rm spon}=26$ ns, and imposing the condition that during the natural lifetime of the excited state, $\tau_{\rm spon}$, the change in the distance between the interacting particles is in the range  $\lambda/50-\lambda/100$, we obtain $p(\Delta v)\approx 0.0011-0.00055$. Then for three components, the probability is $p(\Delta v)^3\approx 1.3\times 10^{-9}-1.67\times 10^{-10}$. Multiplying by the latter value the particle density in the vapor, we obtain $n_{\rm vc}\approx 666152 -82803$ m$^{-3}$, which by \eqref{sm:r} implies $\bar{r}\approx 0.6-1.2$ cm.

\emph{Estimation of the average thermal collision rate of atoms. ---}
The collision rate $\gamma_j$ of a species $j$ in a vapor species $i$ is \cite{Budker.2008}
\begin{equation}
\gamma_j = \sum_i n_i \sigma_{i, j} \bar{v}_{i, j} \, ,
\end{equation}
with particle densities $n_i$, average relative velocities $\bar{v}_{i, j}$, and estimated collisional cross sections $\sigma_{i, j} \approx \pi (r_i, + r_j)^2$, where $r_{i,j}$ are calculated atomic radii \cite{Clementi.1967}.
The relative velocities are given by $\bar{v}_{i, j} = \sqrt{8 k_\mathrm{B} T/\pi \mu_{i, j}}$ with 
the Boltzmann constant $k_\mathrm{B}$, the vapor temperature $T = \qty{22}{\celsius}$, and the reduced masses $\mu_{i, j} = m_i m_j/(m_i + m_j)$ ($m_i$: atomic mass).
The particle densities are estimated from the vapor pressures \cite{Alcock.1984}.
The collision rate of potassium in the vapor cell containing the species K, Rb, and Cs is
\begin{equation}
\gamma_\text{K}
= \sum_{i=\K, \Rb, \Cs} n_i \sigma_{i, \K} \bar{v}_{i, \K} 
\approx \qty{20}{\Hz} \, .
\end{equation}
The different contributions are listed in Tab.~\ref{tab:collisions}.
The average collision time $1 / \gamma_\K \approx \qty{50}{\ms}$ is about 6 orders of magnitude slower than the natural lifetime of the K $4^2\mathrm{P}_{3/2}$ level of \qty{26.37}{\ns} \cite{Wang.1997} and, thus, the influence of collisions on the population redistribution can be neglected.

\begin{table}
\caption{Values used for the estimation of the average interatomic collision time.}
\label{tab:collisions}
\centering
\begin{tabular}{ccccccc}
\hline \hline
species & density \cite{Alcock.1984} & atom radius \cite{Clementi.1967} & coll. cross sec. & rel. velocity & coll. rate & coll. time \\ 
$i$ & $n_i$ (\unit{\per\cubic\m}) & $r_i$ (m) & $\sigma_{i,\K}$ (m$^{2}$) & $\bar{v}_{i, \K}$ (m/s) & $\gamma_{i, \K}$ (Hz) & $1/\gamma_{i, \K}$ (s) \\ 
\hline\\ 
K & \num{5.13e+14} & \num{2.43e-10} & \num{7.42e-19} & \num{565} & \num{0.215} & \num{4.644} \\ 
Rb & \num{1.05e+16} & \num{2.65e-10} & \num{8.11e-19} & \num{483} & \num{4.094} & \num{0.244} \\ 
Cs & \num{3.84e+16} & \num{2.98e-10} & \num{9.19e-19} & \num{455} & \num{16.051} & \num{0.062} \\ 
total & \num{4.94e+16} &  &  &  & 20.361 & 0.049 \\ 
\\
\hline\hline
\end{tabular}
\end{table}



\end{document}